\documentstyle[aps]{revtex}
\input{epsfig.sty}
\newcommand{\beq}{\begin{eqnarray}}
\newcommand{\eeq}{\end{eqnarray}}
\begin{document}
\title{Magnetic Wall From Chiral Phase Transition and CMBR Correlations}
\author{Leonard S. Kisslinger$^\dagger$\\
    Department of Physics, Carnegie Mellon University, Pittsburgh, PA 15213}
\maketitle
\indent
\begin{abstract} Possible CMBR correlations are estimated for a model in 
which a Hubble-size magnetic wall is formed during the QCD chiral phase
transition. Measureable polarization correlations are found for $l$-values
greater than about 1000. It is also found that metric perturbations from 
the wall could give rise to observable CMBR correlations for large $l$.
  
\end{abstract}

\vspace{0.5 in}

\noindent
PACS Indices:98.80.Cq,98.80Hw,12.38.Lg,12.38.Mh \\
\vspace{2 mm}
$^\dagger$ email: kissling@andrew.cmu.edu

\section{Introduction}

\hspace{.5cm}

   In the present paper the possibility is explored that large-scale 
magnetic structures created during the QCD (Quantum Chromodynamic) 
chiral phase transition might lead to observable CMBR  (cosmic microwave 
background radiation) correlations . This work is motivated by improved CMBR
observations in progress, which promise polarization and
temperature correlation measurements with multipoles $l$ in the thousands.
These magnetic structures could also be primoidal seeds of galactic and 
extra-galactic magnetic fields. A long-standing problem of astrophysics 
is the origin of the large-scale galactic and extra-galactic magnetic 
fields which have been observed. Of particular importance for cosmology is 
the possible seeding of these magnetic structures by primoidal, 
early-universe, magnetic structures. For a recent review see Ref\cite{gr}.
We do not, however, investigate galactic or extra-galactic
magnetic structure seeded by the QCD phase transition in the present work, 
but center on CMBR polarization and gravitational wave correlations,

   In most of the theoretical treatments, including inflationary models,
the magnetic fields arise from electrically charged particle motion.
Considerations of nucleosynthesis and CMBR and the galactic magnetic 
fields have been used to constrain the magnitudes of the primoidal tangled
(random) magnetic fields to values of about 10$^{-9}$
Gauss, although in a recent model\cite{dfk} fields of two orders of magnitude
larger are consistent with CMBR observations. Constraints on homogeneous 
primoidal magnetic fields from CMBR have also been determined\cite{bfs}.

Early universe phase transitions are of great interest. The electroweak and 
QCD chiral phase transitions are of particular interest as common ground
for particle physics and astrophysics. For the electroweak phase transition,
using an Abelian Higgs model\cite{kv} with the QED lagrangian included, 
magnetic field generation has been calculated\cite{ae,cst} in bubble 
collisions. For the QCD phase transition primoidal magnetic fields 
generated from charged currents at the bubble surfaces during the nucleation 
have been estimated\cite{co,soj} to be as large as 10$^8$ Gauss, and still
be consistent with observed values of galactic and extra-galactic fields.
There has also been a recent calculation\cite{kl} of the CMBR power spectrum 
from density perturbations caused by promoidal magnetic fields that is
similar to the calculation of Ref\cite{dfk}, but for scalar perturbations.

The present paper is motivated by our QCD instanton model of bubble walls
formed during the QCD chiral phase transition\cite{lsk} and by the recent 
work of Forbes and Zhitnitsky, who have used a QCD domain wall\cite{fz1}
as a mechanism for generating magnetic fields which could evolve to 
large-scale galactic fields\cite{fz2}. Considering classical theory of bubble 
collisions, in which walls with the same surface tension as the colliding 
bubbles are formed within the merged bubbles, it was conjectured\cite{lsk} 
that hubble-scale instanton walls might be formed, with lifetimes sufficient 
to form magnetic walls.
In section II the description of the QCD phase transition in our instanton
model and the possible resulting magnetic wall are discussed. It is shown
that with an interior instanton wall, which is similar to QCD domain walls,
the magnetic wall formed in the hadronic phase is similar to that modelled 
in Ref\cite{fz2}.  In section III the evolution of the magnetic structure
to the time of recombination, and the CMBR polarization correlations are 
derived. In section IV the correlations from metric perturbations are derived,
and in section V we give our conclusions.

\section{QCD Chiral Phase Transition and Magnetic Wall}

\hspace{.5cm}

  In our model\cite{lsk} of the bubbles of the hadronic-phase 
universe nucleating within the quark-gluon phase universe during the QCD
chiral phase transition, which we assume is first order, we use the purely
gluonic QCD Lagrangian density, $ {\cal L}^{glue} = G_{\mu\nu}^a
G^{\mu\nu a}/4$. $G^{\mu\nu} =\partial_\mu A_\nu
 -  \partial_\nu A_\mu -i g [A_\mu,A_\nu]$ is the color field tensor, 
defined in terms of the gluonic color field 
 $A_\mu = A_\mu^n \lambda^n/2$, where $\lambda^n$ are the eight SU(3) 
Gell-Mann matrices, with the notation that $(\mu,\nu)$= (1...4) and
(a,b,...) = (1,2,3) are Dirac indices and color indices, respectively. 
Recognizing that it has been shown that QCD instantons 
can represent the midrange nonperturbative aspects of QCD, we use the 
instanton model\cite{inst} for the color field
\beq
\label{instanton}
 A_\mu^{n,inst}(x)& = & \frac{2 \eta^{-n}_{\mu\nu}x^\nu}{(x^2 + \rho^2)}\\
        G^{n,inst}_{\mu\nu}(x) & = &  -\frac{\eta^{-n}_{\mu\nu} 4 \rho^2}
{(x^2 + \rho^2)^2},
\eeq
for the instanton and a similar expression with -n for the anti-instanton, 
where $\rho$ is the instanton size and the  $\eta^{n}_{\mu\nu}$ are defined
as\cite{inst} 
\beq
\label{eta}
        \eta^a_{ij} & = & \epsilon_{aij} \\ \nonumber
      \eta^a_{\mu4} & = & \delta_{a\mu} \\ \nonumber
      \eta^a_{4\nu} & = & -\delta_{a\nu},
\eeq
with (i,j) = (1,2,3) and $a,\mu,\nu$ as defined above. The QCD instanton
model, reviewed in Ref\cite{ss1}, has been a very useful representation
of nonperturbative QCD for hadronic properties. For example, one can
successfully predict the known scalar glueball candidates using this 
formalism\cite{lsk1}. 

Noting that the wall of the bubble separating the quark-gluon phase
from the hadronic phase is gluonic in nature, and that the instanton model
is successful in representating midrange nonperturative QCD, we have 
recently formulated\cite{lsk} an instanton model of the bubble wall.
The starting point is the energy-momentum tensor of pure gluonic QCD:
\beq
\label{enmom}
T^{\mu \nu} &=&  G^{\mu\alpha}_a G_{\alpha a}^\nu
-\frac{1}{4}g^{\mu \nu} G^{\alpha \beta}_a G_{\alpha \beta a} ,
\eeq
which gives in the instanton model at finite temperature in Minkowski
space the spatial energy momentum tensor
\beq
\label{momentum}
  T^{ij,inst} & = & \, \left( \frac{4 \overline{\rho}^2 \overline{N}}
{(x^2 +\overline{\rho}^2)^2} \right)^2 \, \delta_{ij},
\eeq
and the energy density, 
\beq
\label{T00}
   T^{00,inst} & = &  96 \left( \frac{\overline{\rho}^2 \overline{N}}
{(x^2 +\overline{\rho}^2)^2} \right)^2 ,
\eeq
where $\overline{\rho}$ is the instanton size and $\overline{N}$ is the
instanton density at the bubble surface at temperature T=T$_c$, the
temperature of the chiral phase transition, approximately 150 MeV.
 $\overline{N}$ is determined from the tunneling amplitude\cite{inst,ss1},
and $\overline{\rho}$ has been found in finite temperature 
calculations\cite{chu,ss2} to be  $\overline{\rho} \simeq 0.25 fm$.
In Ref\cite{lsk} the energy density, $T^{0 0}$ was shown to be consistent 
with numerical calculations of the surface tension. $T^{ij}$ can be used 
to calculate bubble collisions.

Recently, effective field models have been used to calculate QCD domain
walls which could form within bubbles during the QCD chiral phase 
transition\cite{fz1}. These domain walls have a space-time structure very 
similar to a wall composed of instantons, with the form given in 
Eq.(\ref{instanton}). Such a wall could interact with nucleons in the 
hadronic phase to produce electromagnetic structures via effective 
magnetic and electric dipole moments\cite{cvvw} of the nucleon field. 
This model was used\cite{fz2} to investigate possible primoidal magnetic 
fields at the time of the chiral phase transition that could lead to 
large-scale galactic magnetic fields.

As pointed out in Ref.\cite{lsk}, the collision of nucleating bubbles
during the phase transition could lead to an interior gluonic wall. If
the theory of classical bubble collisions can applied, the interior wall
would be similar to the bubble walls, with the same surface tension. In
other words, there would be an instanton wall with an energy-momentum tensor
given by Eq.(\ref{momentum},\ref{T00}). In a 1+1 model of colliding bubble 
walls based on QCD we recently found\cite{jck} that, with instanton-like 
boundary conditions, an instanton-like internal wall does seem to form 
in the collision region. Although this model is too simple for studying
the nucleation problem, there have been a number of investigations of the
QCD chiral phase transition using classical nucleation theories with 
effective Lagrangians\cite{ikkl,cm,ssw,is}. These studies find that in
contrast to the electroweak phase transition where many bubles nucleate,
collide and merge, the QCD chiral phase transition seems to proceed via
inhomogeneous nucleation, with larger distance between bubbles and rather
few nucleating bubbles involved in the transition to the hadronic phase.
Therefore, it is likely that very few large-scale instanton walls
were formed, and that the calculation of CMBR correlations from one
magnetic wall, which is the picture used in the present study, is a good
starting point.

 Recognizing that the mathematical form
of the instanton and domain walls are very similar, the arguments of 
Ref.\cite{fz2}, including estimates of the lifetime of the interior QCD
instanton wall, can be applied to estimate the primary electromagnetic wall 
that might have been formed at t$\simeq 10^{-4}$ sec. The magnetic wall is
formed by the interaction of the nucleons with the gluonic wall, with the
electromagnetic interaction Lagrangian
\beq
\label{Lint}
     {\cal L}^{int} & = & -e \bar{\Psi} \gamma^\mu A^{em}_\mu \Psi,
\eeq
where $\Psi$ is the nucleon field operator.
This leads to the electromagnetic interaction with the nucleons magnetic
dipole moment given in terms of the electromagnetic field tensor, 
$F^{\mu\nu}$ by
\beq
\label{Vint}
      {\cal V}^{int} & = & \frac{e}{2 M_n} \bar{\Psi} \sigma_{\mu\nu}
\gamma_5 \Psi F^{\mu\nu},
\eeq
and a similar interaction (without the $\gamma_5$) for the electric dipole
moment, due to cp violation. From Eq.(\ref{Vint}) one can estimate the 
magnetic field in the wall. In analogy to the classical theory in which the 
magnetic field within a magnetized object is given by the magnetic dipole 
moment density\cite{jackson} within a factor depending on the shape of the 
object, the B-field is given by the matrix element of
$\bar{\Psi} \sigma_{\mu\nu} \gamma_5 \Psi$.
 For the gluonic instanton wall oriented in the x-y direction one 
obtains for $B_z \equiv B_W = F^{21}$ within the wall of thickness $\rho$
\beq
\label{bz}
     B_z & \simeq & \frac{1}{\rho\Lambda_{QCD}} \frac{e}{2 M_n} 
 < \bar{\Psi} \sigma_{21}\gamma_5 \Psi > .
\eeq
A suppression factor of $(\rho \Lambda_{QCD})^{-1}$ has been used in
Eq.(\ref{bz}) since alligned dipoles tend to cancel, as discussed at length 
in Ref\cite{fz2}. The matrix element in Eq.(\ref{bz}) is estimated using the
Fermi momentum in the plane of the wall during the QCD phase transition
as $\Lambda_{QCD}$, giving  $< \bar{\Psi} \sigma_{21}\gamma_5 \Psi > = 
\frac{4\pi}{(2\pi)^2} \Lambda_{QCD}^2$, with a factor of four from the spin
and isospin degeneracy. Using $ \Lambda_{QCD}$ = 150 Mev, the resulting 
magnitude of the magnetic field at the wall is
\beq
\label{bw}
     B_W & \simeq & \frac{3 e}{14 \pi} \Lambda_{QCD},
\eeq
which is essentially the same as the estimate of Ref\cite{fz2}, within
the errors of the scale factors. The calculation of the electric field 
is similar, giving $E_z \simeq B_z  \simeq B_W \simeq 10^{17}$ Gauss 
(within the wall). 

Therefore our picture is that at the end of the QCD chiral phase
transition there is a magnetic wall in the hadronic phase, which we
model as
\beq
\label{wall}
  {\bf B}_W({\bf x}) & = & B_W e^{-b^2(x^2 + y^2)} e^{-M_n^2 z^2},
\eeq
or in momentum space
\beq
\label{wallk}
  {\bf B}_W({\bf k}) & = & \frac{B_W}{2\sqrt{2}b^2M_n} e^{-(k_x^2 + k_y^2)
/4b^2} e^{-k_z^2/4M_n^2},
\eeq
where $b^{-1}$ is of the scale of the horizon size, $d_H$, at the end of the 
chiral phase transition (t $\simeq 10^{-4}$ s), $b^{-1} =d_H \simeq$ few km, 
while $M_n^{-1} \simeq 0.2 fm$. 
Therefore, although $B_W$ is very large, since the wall 
occupies a very small volume of the universe, such a structure is 
compatible with nucleosynthesis, galaxy structure and the present CMBR 
observations.

In the present work we conjecture that the QCD bubble collisions lead
to a magnetic wall given by Eq.(\ref{wall}) at $10^{-4}$s and study the 
effects on CMBR polarization correlations and metric fluctuations.

\section{CMBR Polarization Correlations}

\hspace{.5cm}

  In this section we investigate the polarization correlations arising from
the electric and magnetic fields given in Eq.(\ref{wall}). The primoidal
magnetic wall of the present work is quite different from the tangled 
fields considered previously\cite{sb,ss}, however, the evolution to the last 
scattering surface has many features in common with these studies. 
In addition to the polarization anisitropy which results from the magnetic
wall itself, which gives rise to B-type polarization anisitropies discussed
in the next subsection, in the subsequent subsection we also show that 
E-type anisotropies arise from scattering of the background radiation from 
the nucleon moments at the time of the phase transition, and show that
the resulting power spectrum, $C^{EE}_l$, is too small to be detected.

\subsection{Polarization Correlations From Magnetic Wall}

\hspace{.5cm}

  In treating the temperature and polarization matrix for the Stokes
parameters\cite{kks,zs} we use the angular representation of Ref.\cite{hw}
to derive the power spectrum of B-type polarization anisitropies, $C^{BB}_l$,
which arise to a good approximation from the polarization source $P^{(1)}
 = -E^{(1)}_2/\sqrt{6}$, given by
\beq
\label{source}
   E^{(1)}_2 & = & \frac{1}{4}\sqrt{\frac{5}{6\pi}}\int d\Omega 
E(\hat{n},x,\eta)(Y_2^1 - \sqrt{5}Y_1^1),
\eeq
where the electromagnetic wave is propagating with ${\bf k} = k\hat{n}$,
$\eta$ is conformal time, and the $Y_l^m$ are the standard spherical
harmonics.
From our model the source at the wall is $E^{(1)}_2 = 5 B_W/12 \sqrt{2} b^2 
M_n \equiv {\cal B}_W$ at the wall. The solutions of the Bolzman 
equation give the quantities $B_l^{m}$\cite{hw} at the time of recombination
needed for the $C^{BB}_l$. To get the power spectrum we must evaluate 
$<B_z({\bf k},\eta)B_z({\bf k'},\eta)>$. Using the fact that
$exp(-k^2 d_H^2) \simeq 1.0$ at the time of the chiral phase transition,
\beq
\label{bcor}
   <B_z({\bf k},\eta)B_z({\bf k'},\eta)> & \simeq & {\cal B}_W^2
 \delta(k_x-k'_x) \delta(k_y-k'_y)<e^{-k_z^2/4M_n^2}e^{-k^{'2}_z/4M_n^2}> 
\nonumber \\
       & \simeq & {\cal B}_W^2 d_H e^{-k_z^2/4M_n^2} \delta({\bf k-k'}).
\eeq 
This gives for the polarization
power spectrum
\beq
\label{Bpower}
   C^{BB}_l & = & \frac{(l+1)(l+2)}{\pi}{\cal B}_W^2 d_H \int dk k^2
 \frac{j_l^2[k(\Delta \eta)]}{k^2(\Delta \eta)^2},
\eeq
where the conformal time integral over the visibility function has been
carried out and $\Delta \eta$ is the conformal time width at the last
scattering. The integral over the spherical Bessel function is carried
out by using the fact that $j_l(x)$ peaks at $l$ and that the integral
$\int dz j_l^2(z) = \pi/(4l)$ for large $l$. Therefore in the range
$ 100 < l < 2000$ we have the approximate result
\beq
\label{cbb}
    C^{BB}_l & \simeq & \frac{25 d_H^5 B_W^2}{1152 M_n^2 \Delta\eta^3}l^2.
\eeq
Using the parameters $M_n\Delta\eta =1.5 \times 10^{39}$ (from 
Refs.\cite{ss,hs}), $d_H = 0.37\times 10^{24} GeV^{-1}$, and $B_W = 
1.0\times10^{17}$ Gauss, 
\beq
\label{cbbf}
    C^{BB}_l & \simeq & 4.25\times10^{-8} l^2
\eeq

The result for the B-type power spectrum is shown in Fig. 1 by the
solid line.  There have been a number of investigations of the
polarization predicted by inflationary models\cite{dkk} which
show that the B-type polarization in inflationary models is smaller
than the E-type, and that $ l(l+1) C^{BB}_l$ peaks at $l$-values about 100.
Ref\cite{kk} reviews, with extensive references, predictions of inflationary 
models and effects resulting from cosmological phase transitions.  
In Fig. 1 results from Ref\cite{spt} are shown. The curve shown by small
circles is a string model normalized at $l$ = 100 to our magnetic wall
value, and with the same normalization the dashed curve gives a typical
inflationary model result. Even the string model of Ref.\cite{spt} is
two to three orders of magnitude smaller than the T-spectrum at the peak,
One can see that for $l \simeq 1000$ the values of $ C^{BB}_l$ predicted by
our magnetic wall picture exceed those of inflationary and topological 
models, and the $l$ dependence is quite different.
\begin{figure}
\centerline{\psfig{figure=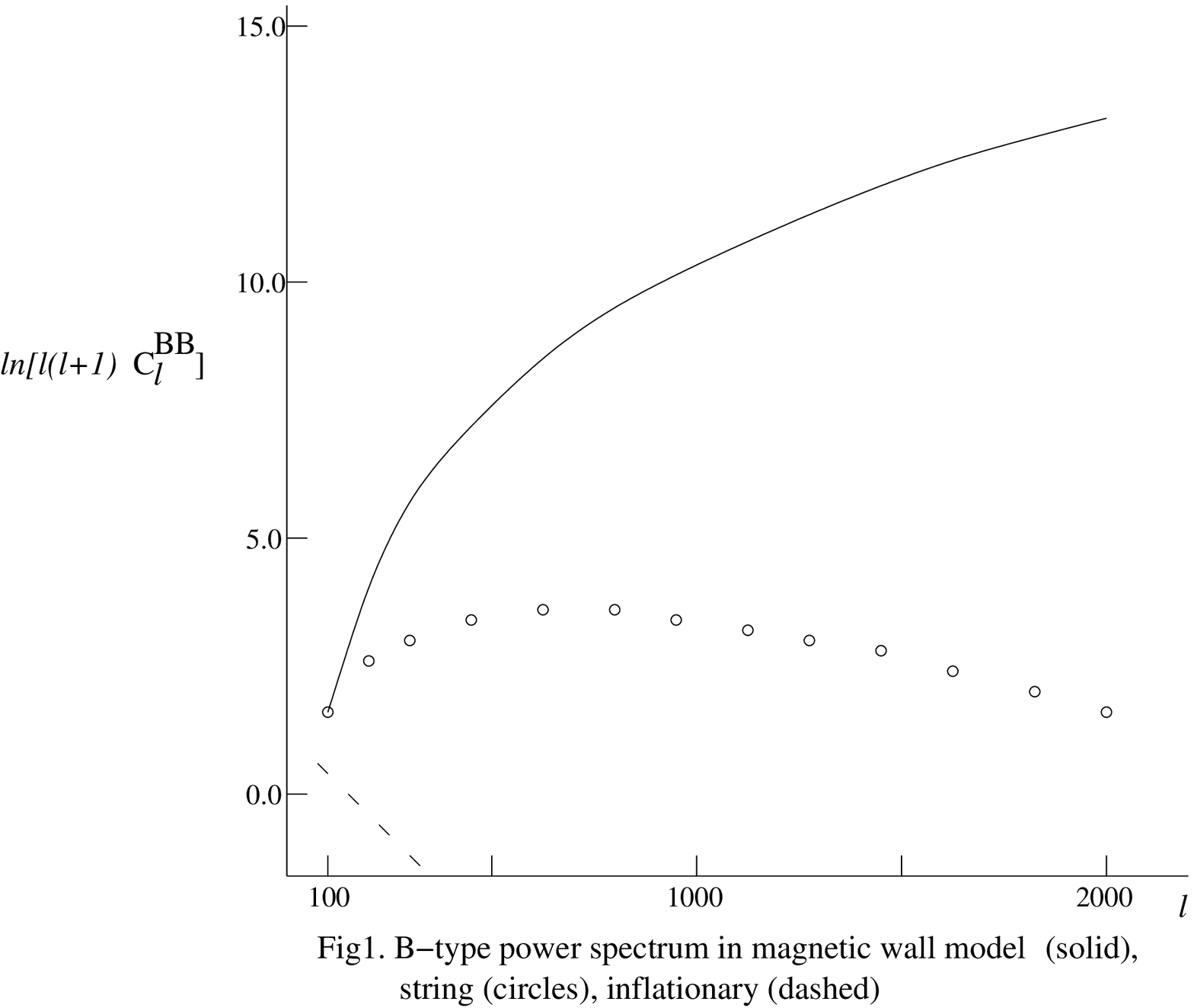,height=7cm,width=12cm}}
{\label{Fig.1}}
\end{figure}

\subsection{Polarization Correlations From Magnetic Dipole Scattering}

\hspace{.5cm}

In this subsection we discuss another possible source of polarization
correlation arising from the horizon-size instanton wall created during
the chiral phase transition and leading to an orientation of nucleon
magnetic dipole moments. As the nucleon moments are alligned along the
wall the background radiation will scatter from these moments producing
scattered waves and magnetic structure which would give polarization at
the last scattering surface. Taking the thickness of the instanton wall
as the scale for the volume associated with the magnetic dipole moments,
The density of moments, n, is obtained from the strength of the magnetic
field:
\beq
\label{bscat}
     B_z & = & \frac{2n\mu_n}{r_W^3},
\eeq
where $\mu_n$ is the neutron magnetic dipole moment and $r_W$ is the wall
thickness. Making use of the torque of the magnetic moment in a magnetic
field ${\bf B}$, $d{\bf m}/dt = \mu_n^2 {\bf s} \times {\bf B}$, with
${\bf s}$ the spin of the neutron, one obtains the magnitude of the scattered
B-field at time t=$10^{-4}$ s: $B_{scat} = n^2 \mu^2/2 cos(\theta_s) k E_0
exp(ikr)/r$, with $\theta_s$ the angle between the neutron spin and the
direction of the incident radiation and $ E_0$ the E-field of the incident
radiation. From this one obtains the Stokes parameter U, and in the notation
of Ref\cite{hw} the polarization source
\beq
\label{b2scat}
   B_2^{0}({\bf k},\eta) & = & \frac{k^2 E_0^2}{2\sqrt{6}}(n\mu_n^2)^2.
\eeq
Note that $ E_2^{0}({\bf k},\eta) =  B_2^{0}({\bf k},\eta)$, and results
in E-type polarization anisotropies, $C^{EE}_l$.

  Without a detailed calculation one sees that the resulting polarization
correlations will be very small, since the extra $k^2$ dependence from the
scattering from the moments introduces a factor of $ (l^2/\Delta\eta)^2$.
Therefore the $C^{EE}_l$ polarization anisotropies resulting from the
scattering from the nucleon moments at the time of the phase transition are 
too small to be measured.
 
\section{Metric Perturbations From Magnetic Wall and CMBR Power Spectrum}

\hspace{.5cm}

  In this section we derive the power spectrum from the gravitational waves
arising from the magnetic wall of Sec.2. The calculation is very similar
to that of Ref.\cite{dfk}, in which the power spectra was derived
for tangled magnetic fields such as those considered in Refs.\cite{sb,ss}, 
with various scenarios for the scale dependence. Although our model of the 
narrow hubble-size magnetic structure is quite different, we can make use 
of much of the formalism of Ref.\cite{dfk}.

  The stress-energy tensor\cite{jackson} for our model magnetic wall, which
has only the 33 spacial component, is
\beq
\label{t33}
   T_{33}({\bf k}) & = & \frac{1}{8\pi} \int d^3q B_3({\bf q})B_3({\bf k-q})
 \\ \nonumber
          & = & \frac{2\pi^3\sqrt{\pi}}{M_n}B_W^2 e^{-\frac{3k_3^2}{8M_n^2}}
\eeq
at the time of the chiral phase transition. Note that we are assuming that
the wall is of hubble size in the x-y directions and have omitted the
parts of the expression for the $k_x,k_y$ dependence shown in Eq.(\ref{wallk}).
From Eq.(\ref{t33}) we obtain the B-wall power spectrum
\beq
\label{bpower}
  < T_{33}({\bf k},\eta) T_{33}({\bf k'},\eta)> & = & \int d^3q d^3q'
 < B_3({\bf q}) B_3({\bf k-q}) B_3({\bf q'}) B_3({\bf q}) B_3({\bf k'-q'})>.
\eeq
In evaluating Eq.(\ref{bpower}) we use
\beq
\label{length}
    < e^{-\frac{k^2}{8M_n^2}} e^{-\frac{k^{'2}}{8M_n^2}}\delta(k_x)\delta(k'_x)
 \delta(k_y)\delta(k'_y) & = &  e^{-\frac{k^{'2}}{4M_n^2}}d_H 
 \delta({\bf k - k')}).
\eeq
Using the notation of Ref\cite{dfk} with a tensor projection,
\beq
\label{bpower1}
     < T_{33}({\bf k},\eta) T_{33}({\bf k'},\eta)> & = & \frac{4}{a^8} 
  f^2(k^2) \delta({\bf k - k')}),
\eeq
with
\beq
\label{f}
    f^2(k^2) & = & \frac{2^3 \pi^9}{M_n^2}d_H B_W^4.
\eeq
With $h_{ij}$ the tensorial perturbations of the Friedman universe,
$ds^2 = a^2[-d\eta^2 +(\delta_{ij} + 2h_{ij})dx^idx^j]$. The Einstein 
equations, using the representation  $h_{ij}= 2HQ^{(2)}_{ij}$ of 
Ref\cite{hw}, with $\nabla^2Q^{(2)}_{ij}=-k^2Q^{(2)}_{ij}$  are
\beq
\label{einstein}
   \ddot{H}+2\frac{\dot{a}}{a}  \dot{H}+k^2H & = & 8\pi G \frac{4}{a^8}
  f^2(k^2).
\eeq
The power spectrum for the tensor metric fluctuations are given by
\beq
\label{mpower}
 <\dot{h}_{ij}({\bf k'},\eta)\dot{h}_{ij}({\bf k},\eta)> & = & 
4|\dot{H}({\bf k)})|^2 \delta({\bf k - k')}).
\eeq
Since Eq(\ref{einstein}) was investigated in Ref\cite{dfk}, except with a
different magnetic stress tensor, we use the solutions that they obtained.
Assuming that the magnetic wall is formed at t=$10^{-4}$ s in the radiation
dominated epoch, at redshift $z_{in}$, and neglecting perturbations created
after the time of matter-radiation equilibration $\eta_{eq}$, for 
$\eta > \eta_{eq}$ the approximate solution for $\dot{H}$ is
\beq
 \label{H}
   \dot{H}(k,\eta) & = & 4\pi G \eta_o^2 z_{eq}ln(\frac{z_{in}}{z_{eq}})
 \frac{j_2(k\eta)}{\eta}f(k).
\eeq
Carrying out integrals over Bessel functions, the solution
for the metric fluctuation power spectrum is for $l>>1$
\beq
\label{mcl}
  C_l & = & \left[\frac{14}{25}Gz_{eq} ln\frac{z_{in}}{z_{eq}}\right]^2 l^5 
 \eta_o^2 \int dz \frac{1}{z^4}f^2(z/\eta_o)J_{l+3}(z).
\eeq
Using the form of f(z) given in Eq(\ref{f}), the integral in Eq(\ref{mcl})
is approximately
\beq
 \label{integral}
     \int dz \frac{1}{z^4}f^2(z/\eta_o)J_{l+3}(z) & \simeq & 
 \frac{8 \pi^9 d_H}{M_n^2}B_W^4\frac{0.106}{l^4}
\eeq
for $M_n \eta_o >> l >> 1$. Taking $t_{in}$ at $10^{-4}$ s or 
$z_{in} = 1.17 \times 10^8 z_{eq}$ and $d_H = 0.371 \times 10^{24} GeV^{-1}$,
the power spectrum is ($l >> 1$)
\beq
\label{power}
 l(l+1)C_l & \simeq & 6.9 \times 10^{-7} l^3
\eeq
for $B_W = 10^{17}$ Gauss. Therefore, the metric perturbations from the
QCD-induced wall result in CMBR effects large compared to other tensor 
perturbations that have been estimated for for $l$ values of the order
of 1000.

\section{Conclusions}

\hspace{.5cm}

  We conlcude that if the QCD chiral phase transition produces a gluonic
wall of domain size and nucleon thickness at the time of the final
collision of nucleating bubbles, and produces a magnetic wall as that
found in the domain wall model of Refs\cite{fz1,fz2}, it would result
in polarization correlations which have a $l$-dependence for $l \simeq 1000$
different than other cosmological predictions and which should be measurable 
with the next generation of CMBR measurements. Also, there would be 
distinguishable temperature correlations arising from metric fluctuations.
In order to investigate the collisions during nucleation to determine the 
details of the interior wall it is necessary
to include quark and hadronic degrees of freedom to obtain the difference
in the free energy in the two phases, a subject for future research.
\vspace{2mm}

\centerline{\bf Acknowledgements}
\bigskip
The author would like to acknowledge helpful discussions with Ariel
Zhitnitsky on his domain wall model, and helpful discussions with
Ho-Meoyng Choi, Ernest Henley, Pauchy Hwang and Mikkel Johnson.
This work was supported in part by NSF grant PHY-00070888 and by the 
Taiwan CosPA Project, Taiwan Ministry of Education 89-N-FA01-1-3,
and in part by the DOE's Institute of Nuclear Theory
at the University of Washington while the author was in residence.

\end{document}